\documentclass[aps,prd,preprint,superscriptaddress,nofootinbib]{revtex4-1}
\usepackage{amsmath,amssymb,mathrsfs,mathtools,hyperref,xcolor}

\def\my{\mathfrak{y}}

\def\ep{\epsilon}
\def\cO{{\cal O}}

\def\ra{\rightarrow}

\newcommand{\be}{\begin{equation}}
\newcommand{\ee}{\end{equation}}
\def\beq#1\eeq{\begin{align}#1\end{align}}
\newcommand{\nn}{\nonumber}

\begin{document}
\title{Re-visiting Supersymmetric Janus Solutions: \\A Perturbative Construction}
\author{Nakwoo Kim}
\email{nkim@khu.ac.kr}
\affiliation{Department of Physics and Research Institute of Basic Science,
	Kyung Hee University, Seoul 02447, Republic of Korea}
\affiliation{School of Physics, Korea Institute for Advanced Study, Seoul 02445, Republic of Korea}
\author{Se-Jin Kim}
\email{power817@khu.ac.kr}
\affiliation{Department of Physics and Research Institute of Basic Science,
	Kyung Hee University, Seoul 02447, Republic of Korea}
%\date{\today}

\begin{abstract}
We construct holographic Janus solutions, which describe a conformal interface in the theory of M2-branes, in four-dimensional gauged supergravities using a perturbative method. In particular, we study three Einstein-scalar systems and their BPS equations, which are derived by Bobev, Pilch, and Warner (2014) \cite{Bobev:2013yra}. The actions of our interest are all consistent truncations of $D=11$ supergravity chosen to be invariant under $SO(4)\times SO(4)$, $SU(3)\times U(1)\times U(1)$, and $G_2$ symmetry subgroups of $SO(8)$ respectively. The utility of our semi-analytic result is illustrated by the calculation of minimal area surface and the associated holographic entanglement entropy.
\end{abstract}

\maketitle
\newpage
\section{\label{sec:1}Introduction}
Conformal field theories constitute an important subset of quantum field theories thanks to their extended symmetry algebra, which includes in particular the scale transformation. Because of scale invariance, conformal field theories are crucial in the study of critical phenomena near phase transition. Another reason why there has been so much interest on conformal field theory over the past decades is the holographic principle, in particular the AdS/CFT correspondence \cite{Maldacena:1997re}. According to it, a strongly-coupled conformal field theory with a large number of degrees of freedom can have a dual description as a weakly-coupled Einstein-like 
gravity in one higher spacetime dimensions\footnote{For more careful discussion on the requirement for the conformal field theory to have a gravity dual, see {\it e.g.} \cite{Heemskerk:2009pn,ElShowk:2011ag}.}. Operators in conformal field theory have their dual fields in the gravity counterpart, and an elaborate prescription for calculation of correlation functions thereof has been established and passed a number of non-trivial tests \cite{Aharony:1999ti}.

One then tries to turn on some deformation in the duality pairs to break the scale invariance and see if the correspondence still holds. Janus configuration \cite{Bak:2003jk,Clark:2004sb} is one of the most interesting examples, where we select a relevant operator and make the dual field in AdS side position-dependent.
Typically we introduce a co-dimension one defect, or interface, and having different values for scalars on each side implies that some of the coupling constants jump across the interface. For the first example considered in \cite{Bak:2003jk}, on the gauge theory side we have ${\cal N}=4$ super Yang-Mills theory in 3+1 dimensions, and across the 2+1 dimensional interface the gauge coupling takes different values. 
On the gravity side, we have to consider so-called domain wall-like solutions, and due to the inherent nonlinearity of Einstein gravity the field equations are typically reduced to a system of non-linear
ordinary differential equations. To obtain an exact solution is thus usually not possible because of nonlinearity. 
Indeed, most
of the previous works on the construction of Janus solutions on gravity side have relied on 
numerical integration \cite{DeWolfe:2001pq,Clark:2004sb,Clark:2005te,DHoker:2006vfr,DHoker:2007zhm,Suh:2011xc,Bobev:2013yra,Bachas:2013vza,Janik:2015oja,Pilch:2015dwa,Bak:2016rpn,Karndumri:2016tpf,Karndumri:2017bqi,Gutperle:2017nwo,Suh:2018nmp,Gutperle:2018fea}. 

Recently we have proposed a new perturbative approach for similar systems of non-linear 
ordinary differential equations derived from Einstein gravity coupled to scalar fields, in the context of 
AdS/CFT correspondence \cite{Kim:2019feb}. This technique
was successfully applied to several Einstein-scalar systems in Euclidean signature \cite{Kim:2019feb,Kim:2019rwd,Kim:2019ewv} which describe mass deformations of several dual conformal field theories in large-$N$ limit \cite{Freedman:2013ryh,Bobev:2013cja,Bobev:2016nua,Gutperle:2018axv,Bobev:2018wbt}. In particular, the matching of sphere partition function for ${\cal N}=2^*$ mass-deformed super Yang-Mills and its supergravity dual is now more firmly established through exact evaluation of some leading expansion coefficients \cite{Kim:2019feb}. For ${\cal N}=1^*$ deformation, the first non-trivial coefficients in the series of expansion of the sphere partition function is analytically computed \cite{Kim:2019rwd}. For the duality proposal of mass-deformed Brandhuber-Oz theory \cite{Brandhuber:1999np}, we managed to re-sum the series expansion form of the sphere partition function as a function 
of mass and argued the result does not agree with the large-$N$ limit of the field theory side computation \cite{Kim:2019feb}.
The main goal of this paper is to illustrate that the same technique can be also successfully applied 
to holographic Janus solutions. Using our semi-analytic solutions, we calculate the holographic 
entanglement entropy \cite{Ryu:2006bv,Ryu:2006ef} as a function of the perturbation parameter which controls the magnitude of the deformation away from the AdS vacuum. 

Let us explain the setup of our interest in more detail. We will consider, for concreteness the 
Janus solutions in three consistently truncated Einstein-scalar systems from ${\cal N}=8,\, D=4$ maximal supergravity with $SO(8)$ gauge group \cite{deWit:1982bul}. The dual field theory is 
the well-known Chern-Simons matter theory living on M2-branes, the action of which was first
explicitly written down by Aharony, Bergman, Jafferis and Maldacena (ABJM) in \cite{Aharony:2008ug}. 
Instead of the full $SO(8)$ gauged supergravity we are interested in various truncated models, focusing on specific
mass deformations.
Such truncated supergravity models were constructed and analyzed in \cite{Bobev:2013yra}, which we closely follow and consider the BPS equations presented thereof. 
We are only concerned about the bosonic sector since we are after classical solutions. All three models have a single complex scalar field which is invariant under a certain subgroup of the global symmetry $SO(8)\subset E_{7(7)}$. They will be referred to as $SO(4)\times SO(4)$, $SU(3)\times U(1)\times U(1)$, and $G_2$ models. The scalar fields, although they will be always called $z$ to maintain the generality of the discussion, are dual to different mass terms in the ABJM theory which preserve different symmetry subgroup of $SO(8)$ which is the R-symmetry of the dual supersymmetric field theory. 
We are interested in conformal defects, which means, the Lorentz symmetry along the defect is also promoted to conformal symmetry and our gravity ansatz is AdS${}_3$-sliced, instead of the Minkowski space. We treat the scalar fields as perturbation and solve the field equations exactly at each order. The boundary condition we impose is that the solution 
should be asymptotically AdS${}_4$ in UV, and regular in IR.

The utility of the perturbative approach is best illustrated when holographic calculations are compared to the field theory side result using supersymmetric localization \cite{Pestun:2016zxk}, where we take the Euclidean signature and put the theory typically on the sphere. 
It was why the BPS equations in \cite{Freedman:2013ryh,Bobev:2013cja,Bobev:2016nua,Gutperle:2018axv,Bobev:2018wbt} were obtained in Euclidean signature with sphere-sliced metric ansatz. Janus solutions in holography are constructed in 
Lorentzian signature on the contrary, so it is not clear to us whether we can compare the result to a localization 
computation result. We thus choose to calculate holographic entanglement entropy which is the minimal area of a spatial surface \cite{Ryu:2006bv,Ryu:2006ef}. Although we do not try to do the field theory side computation in this article, 
we believe it should be possible, at least in weakly-coupled regime and simple geometry of the entanglement region, using {\it e.g.} the replica trick \cite{Calabrese:2004eu} and explicit form of the ABJM action.

Our plan is as follows. In Sec.\ref{sec:2} we setup the notation and present the Einstein-scalar actions and their associated BPS equations we will study. Sec.\ref{sec:3} is the main part where we solve the BPS equations treating scalar fields as perturbation to AdS vacuum. We also consider backreaction and higher orders solutions with the right boundary condition are also obtained analytically. In Sec.\ref{sec:4} we calculate the holographic entanglement entropy for
Janus solutions constructed in Sec.\ref{sec:3}, again solving the minimal-surface condition perturbatively. 
We conclude in Sec.\ref{sec:4} with discussions.

\section{\label{sec:2}Actions and BPS equations}
In this section we closely follow and summarize the setup of \cite{Bobev:2013yra}, as a preparation for our perturbative analysis which will be presented in the next section.
The authors of \cite{Bobev:2013yra} presented three distinct subsectors of ${\cal N}=8$, $SO(8)$-gauged 
supergravity in $D=4$, by requiring invariance under certain symmetry subgroups of the global symmetry $E_{7(7)}$. They all have
a complex scalar field coupled to Einstein gravity, and schematically share the following form.
\beq
e^{-1}{\cal L}=\frac{1}{2} R - {\cal K}_{z \bar z} z' {\bar z}' - g^2 {\cal P}(z,\bar z).
\label{ESaction}
\eeq
In the above $e$ denotes the Jacobian determinant of the metric tensor, $g$ is the gauging parameter {\it i.e.} coupling constant, and ${\cal P}$ is the scalar potential. The actions enjoy ${\cal N}=2$ supergravity structure in four dimensions when the fermionic sector is added appropriately, and the complex scalar $z$ with conjugate $\bar z$ parameterize a K\"ahler manifold $SL(2,\mathbb{R})/SO(2)$, with a K\"ahler potential 
\beq
{\cal K} = - k \log (1-z\bar z) . 
\eeq
The metric in the internal space is as usual calculated by ${\cal K}_{z\bar z} =\partial_z \partial_{\bar z} {\cal K}=k/(1-z \bar z)^2$ and $k$ is a constant which represents in what representation $SL(2,\mathbb{R})$ is embedded inside the larger symmetry group $E_{7(7)}$ of ${\cal N}=8$ gauged supergravity in four-dimensions. On the other hand, the scalar potential is given in terms of holomorphic superpotential ${\cal V}(z)$,
\beq
{\cal P}&=e^{\cal K}({\cal K}^{z \bar z} \nabla_z {\cal V} \nabla_{\bar z} \bar {\cal V} -3 {\cal V} \bar {\cal V})
=4 {\cal K}^{z \bar z}\partial_z W\partial_{\bar z} W -3 W^2 ,
\eeq
where $W\equiv \sqrt{e^{\cal K} {\cal V} \bar {\cal V}}$. In \cite{Bobev:2013yra} the authors considered the dual of a specific linear combination of mass terms on the gauge field theory side, preserving $SO(4)\times SO(4)$, $SU(3)\times U(1)\times U(1)$, and $G_2$ symmetry respectively. Although we use the same symbol, one should keep in mind that $z,\bar{z}$ are thus dual to different mass terms in the dual field theory. For each model, the essential information is given in the table below.

\begin{table}[h!]
\centering
 \setlength\tabcolsep{6.5pt}
 \begin{tabular}{|c|c|c|c|} 
 \hline
 & ${ SO(4)\times SO(4)}$ &${ SU(3)\times U(1) \times U(1)}$ &$ {G_2}$\\ [0.5ex] 
 \hline
 $k$ & 1 & 3 & 7 \\ [0.5ex] \hline
$ {\cal V}/\sqrt{2} $ & $1$ & $ z^3+1$ & $z^7 +7 z^4+7 z^3+1$ \\ [0.5ex] 
 \hline
 \end{tabular}
\end{table}

We now turn to the metric ansatz and the associated BPS equations. Physically speaking we are interested in co-dimension one conformal interfaces, so the spacetime is required to include $AdS_3$. We choose the following metric ansatz:
\beq
ds^2_4 = d\mu^2 +e^{2A(\mu)} \, ds^2(AdS_{3}) ,
\label{metric}
\eeq
where $ds^2(AdS_3)=dr^2-\cosh^2(r/ \ell)dt^2+\sinh^2(r/ \ell) \ell^2 d\phi^2$, with curvature radius $\ell$.
When $e^{2A}=(L/\ell)^2\cosh^2(\mu/L)$ the above metric becomes  exactly $AdS_4$ with curvature radius $L$.

Using the standard parametrization $z:=e^{i\zeta}\tanh\alpha$, one can easily verify that the field equations of \eqref{ESaction} are reduced to the following one-dimensional action:
\beq
{\cal L}&= e^{3A} \left[3 (A')^2 -k\left[(\alpha')^2+\frac{1}{4} \sinh^2(2 \alpha) (\zeta')^2\right]-g^2 {\cal P} \right]-\frac{3}{\ell^2}e^{A},
\eeq
where $(\bullet)'={d(\bullet)}/{d\mu}$.
The scalar potential can be written in terms of superpotential $W$
\be
{\cal P} =\frac{1}{k} \left[\left(\frac{\partial W}{\partial \alpha}\right)^2 +\frac{4}{\sinh^2 (2 \alpha)}\left(\frac{\partial W}{\partial \zeta}\right)^2 \right]-3 W^2 . 
\ee

One can also substitute the ansatz into the Killing spinor equations and demand existence of non-trivial solutions. The analysis of \cite{Bobev:2013yra} concludes that the following first-order differential relations, 
\beq
\alpha' %&= -\frac{1}{k} \left(\frac{A'}{W}\right)\frac{\partial W}{\partial \alpha}+ \frac{2\kappa}{k}\left(\frac{e^{-A}}{l}\right) \frac{1}{\sinh (2 \alpha)}\frac{1}{W} \frac{\partial W } {\partial \zeta},
&= -\frac{1}{2k} \left(\frac{A'}{W^2}\right)\frac{\partial W^2}{\partial \alpha}+ \frac{\kappa}{k}\left(\frac{e^{-A}}{\ell}\right) \frac{1}{\sinh (2 \alpha)}\frac{1}{W^2} \frac{\partial W ^2} {\partial \zeta},
\\
\zeta' %&= -\frac{4}{k} \left(\frac{A'}{W}\right)\frac{1}{\sinh^2 (2\alpha)}\frac{\partial W}{\partial \zeta}- \frac{2\kappa}{k}\left(\frac{e^{-A}}{l}\right) \frac{1}{\sinh (2 \alpha)}\frac{1}{W} \frac{\partial W } {\partial \alpha},
&= -\frac{2}{k} \left(\frac{A'}{W^2}\right)\frac{1}{\sinh^2 (2\alpha)}\frac{\partial W^2}{\partial \zeta}- \frac{\kappa}{k}\left(\frac{e^{-A}}{\ell}\right) \frac{1}{\sinh (2 \alpha)}\frac{1}{W^2} \frac{\partial W^2 } {\partial \alpha},
\eeq
are sufficient for supersymmetry and the field equations to be satisfied when combined with a constraint 
\be
\label{Hcon}
(A')^2=g^2 W^2-\ell^{-2}e^{-2A} .
\ee
For AdS${}_4$ vacuum the scalar fields $\alpha,\zeta$ vanish and their field equations are trivially satisfied. On the other hand \eqref{Hcon} is satisfied for $e^{2A}=(L/\ell)^2\cosh^2(\mu/L)$ where $L$ is related to the vacuum value of $W=\sqrt{2}$ and $L^{-1}=\sqrt{2}g$. $\kappa=\pm 1$ is associated with the choice of Killing spinor projection rule, and we take $\kappa =-1$ for concreteness.

Our strategy is, as illustrated in \cite{Kim:2019feb,Kim:2019rwd,Kim:2019ewv}, to tackle the BPS equations perturbatively. The AdS vacuum is treated as a reference solution {\it at zeroth order}, and scalar excitations will be treated as small perturbations at first, and their backreaction to the metric as well as their self-interaction will be studied iteratively order-by-order in the perturbation parameter. There is a subtlety though. It turns out that the phase part of the scalar, $\zeta$, can be given a non-trivial kink-like profile already at zeroth order. Since the modulus $\alpha$ will be kept zero at zeroth order, this does not make the entire complex scalar $z$ non-vanishing, but this type of zeroth order deformation is essential for non-trivial Janus-like solutions. 

Additionally, it is also worth mentioning here that there will be in general three integration constants we can turn on for the single-scalar models of our interest here. Among them, what is most crucial is the one which corresponds to the strength of the perturbation, while the remaining two are the location of the center of Janus in the spacetime and the internal space. This property is to be contrasted with the supergravity solutions for mass-deformed partition function \cite{Kim:2019feb,Kim:2019rwd,Kim:2019ewv}, where each integration constant is dual to a mass parameter on the field theory side.
%%%%%%%%%%%%%%%%

%%%%%%%%
\section{\label{sec:3}Perturbative solutions}
%%%%%%%%%%%%%
\subsection{$SO(4) \times SO(4)$}
In this case $k=1$ and it turns out that one can integrate the BPS equations exactly. Thus this example serves as a touchstone for the utility of our proposed method, just like the perturbative re-construction of exact solutions \cite{Freedman:2013ryh} in the holographic mass-deformed ABJM theory \cite{Kim:2019feb}. The scalar potential and the superpotential are
\beq
{\cal P}=-2 (\cosh 2 \alpha +2),
%\eeq
%\beq
\quad
{\cal W} = \sqrt{\frac{2}{1-|z|^2}} . 
\eeq
The action allows a conserved Noether charge, because it is independent of $\zeta$. 
\beq
Q = e^{3A}\sinh^2(2 \alpha) \zeta'=const. 
\eeq
We will be able to express this integration constant in terms of the perturbative parameter.
According to the analysis of \cite{Bobev:2013yra}, the BPS equations are, in addition to the universal constraint
\eqref{Hcon},
\beq
\alpha '&=-\tanh \alpha A',
\\
\zeta '&=\frac{e^{-A} }{\ell}\text{sech}^2\alpha .
\label{so4eqs}
%\\
%\left(A'\right)^2=&-\frac{e^{-2 A}}{l^2}+2 g^2 \cosh ^2\alpha .
%\nn
\eeq
They are easily integrated \cite{Bobev:2013yra}, and the branch of solutions which include the AdS vacuum take the following form. 
\beq
e^{A(\mu)} &= \frac{\sqrt{1-a^2}}{\sqrt{2} g \ell} \cosh(\sqrt{2}g(\mu-\mu_{IR})),
\nn\\
\sinh \alpha &=\frac{a}{\sqrt{1-a^2}}\frac{1}{\cosh(\sqrt{2} g(\mu-\mu_{IR}))}, 
\label{esol}
\\
\tan (\zeta-\zeta_{IR})&= \sqrt{1-a^2}\sinh (\sqrt{2} g (\mu-\mu_{IR})).
\nn
\eeq
Here we have three integration constants, as already mentioned: $a<1$ represents the strength of the deformation, and $\zeta_{IR},\mu_{IR}$ denote the location of the defect in internal and external spaces.

Let us illustrate how these solutions can be constructed perturbatively. Using the form of AdS vacuum and also from the consideration of scalar fluctuation equations, we find that the following expansion in $\epsilon$ is most convenient.
\beq
\alpha(\mu) &=\text{sech} \left(\frac{\mu}{L}\right) \sum_{{\rm odd}\, n\ge 1} \alpha_n(\mu)\epsilon^n,
\nn \\
\zeta(\mu) &= \sum_{{\rm even}\, n\ge 0} \zeta_n (\mu)\epsilon^n ,
\label{ansatz1}
 \\
e^{A(\mu)} &= \frac{L}{\ell}\cosh\left(\frac{\mu}{L}\right)\left(1+\sum_{{\rm even}\, n\ge 2} {\cal A}_n(\mu)\epsilon^n\right). \nn
\eeq
Restriction to odd/even powers is possible thanks to the invariance under $\alpha\ra -\alpha$, and when one utilizes the re-parametrization freedom of $\epsilon$. This implies $e^A,\alpha$ are even functions in $\mu$. In the above ansatz we restrict to the case where the position of the Janus defect $\mu_{IR}$ is small: we will see shortly that $\mu_{IR}\sim {\cal O}(\ep^2)$ can be included. We substitute \eqref{ansatz1} into the BPS equations and demand they are satisfied for
all $\ep$. 

Equating zeroth order terms in $\ep$, we have 
\beq
\zeta_0(\mu)&=\zeta_*+\tan ^{-1}\left(\sinh \left(\frac{\mu}{L}\right)\right),
%\nn\\ L&= (\sqrt{2} g)^{-1},
\eeq
where $\zeta_*$ is an arbitrary real number which gives the value of $\zeta$ at IR. 
At first order in $\ep$ the equation for $\alpha_1$ is, up to rescaling of $\ep$, 
\beq
\alpha_1(\mu)&=1 .
\eeq
%where $\alpha_{(1)}$ is an integrate constant and can be setting to $\alpha_{(1)} =1$ by absolving into $\epsilon$.
Then at 2nd order in $\ep$, we have
\beq
\zeta_2(\mu)&=\zeta_{(2)}-\mu_{(2)}\text{sech}\left(\frac{\mu}{L}\right) -\frac{1}{2} \tanh \left(\frac{\mu}{L}\right) \text{sech}\left(\frac{\mu}{L}\right),
\\
{\cal A}_2(\mu) &= -\frac{1}{2} - \mu_{(2)} \tanh\left(\frac{\mu}{L}\right),
\eeq
where $\zeta_{(2)}$, $\mu_{(2)}$ are integral constants. The constant $\mu_{(2)}$ is related to the integration constant in \eqref{esol} via $\mu_{IR} = \mu_{(2)} \epsilon^2$. On the other hand $\zeta_{(2)}$ can be absorbed into $\zeta_*$. The third order solution for $\alpha$ is
\beq
\alpha_3(\mu)&= \alpha_{(3)}+\frac{\mu_{(2)}}{2} - \frac{1}{6} \text{sech}^2\left(\frac{\mu}{L}\right).
\eeq
Since $\alpha_{(3)}$ is the homogeneous solution, we can freely choose its value and we set it to zero. This way we adopt a relation $\ep=a/\sqrt{1-a^2}$. Continuing this way and demanding higher-order solutions decay faster in the UV than lower-order solutions, we can reproduce the solutions in \eqref{esol}. In particular, one can check that $e^{A}$ is just a constant ($\sqrt{1-a^2}=1/\sqrt{1+\ep^2}$) times zeroth order solution (except for shift by $\mu_{IR}$).
%\beq
%e^{A}&=\frac{L}{l}\cosh\left(\frac{\mu}{L}\right)\left(1-\frac{1}{2}\epsilon^2-\frac{1}{8}\epsilon ^4-\frac{1}{16}\epsilon ^6-\frac{5 }{128}\epsilon ^8+{\cal O}(\epsilon^{10})\right)
%\eeq
%This is our result by using perturbation method. 
We also verify that the Noether charge is indeed constant, and consistent with 
\beq
\zeta'e^{3A}\sinh^22\alpha &= 2a^2 g^{-2}\ell^{-3}.
\eeq
up to ${\cal O}(\ep^{10})$.
%From our result this quantity is $4 L^2 a^2/l^2$.

%%%%%%
\subsection{The $SU(3)\times U(1) \times U(1)$ case}
In this case the scalar potential and the real superpotential are given as follows,
\beq
{\cal P}=-6 \cosh 2 \alpha,  \quad {\cal W} =\frac{\sqrt{2} (z^3+1) }{(1-|z|^2)^{3/2}} , 
\eeq
with $k=3$. Since ${\cal P}$ is independent of $\zeta$ here, we have a Noether charge.
\beq Q = \zeta'e^{3A}\sinh^2 2\alpha . \eeq 
The BPS equations are, in addition to the universal constraint \eqref{Hcon}, 
\beq
\alpha '%&=-\frac{4 \sinh \alpha \cosh \alpha A' (\sinh 4\alpha \cos 3\zeta+\cosh 4\alpha+3)}{4 \sinh ^32\alpha \cos 3\zeta+15 \cosh 2\alpha+\cosh 6\alpha}
%\nn\\
%&+\frac{4 e^{-A} \sinh ^22\alpha \sin 3\zeta}{l \left(4 \sinh ^32\alpha \cos 3\zeta+15 \cosh 2\alpha+\cosh 6\alpha\right)},
%\nn\\
&=-\frac{\sinh \alpha \cosh \alpha  (\sinh 4\alpha \cos 3\zeta+\cosh 4\alpha+3)}{2 W^2}A'+\frac{ e^{-A} \sinh ^22\alpha \sin 3\zeta}{2 \ell W^2},
\\
\zeta '%&=\frac{8 \sinh 2\alpha A' \sin 3\zeta}{4 \sinh ^32\alpha \cos 3\zeta+15 \cosh 2\alpha+\cosh 6\alpha}
%\nn\\
%&+\frac{4 e^{-A} (\sinh 4\alpha \cos 3\zeta+\cosh 4\alpha+3)}{l \left(4 \sinh ^32\alpha \cos 3\zeta+15 \cosh 2\alpha+\cosh 6\alpha\right)},
%\nn\\
&=\frac{ \sinh 2\alpha  \sin 3\zeta}{W^2}A'+\frac{ e^{-A} (\sinh 4\alpha \cos 3\zeta+\cosh 4\alpha+3)}{ 2 \ell W^2 },
%\nn\\
%\left(A'\right)^2&=-\frac{e^{-2 A}}{l^2}+g^2 W^2,
%\frac{1}{8}  \left(4 \sinh ^32\alpha \cos 3\zeta+15 \cosh 2\alpha+\cosh 6\alpha\right).
\eeq
We note that the real superpotential $W$ is given as
\beq
W^2=%\frac{1}{8}
\left(4 \sinh ^32\alpha \cos 3\zeta+15 \cosh 2\alpha+\cosh 6\alpha\right)/8.
\eeq

Unlike the $SO(4)\times SO(4)$-symmetric case, these  equations are hard to solve exactly. Since $W^2$ is not even under $\alpha\ra -\alpha$ it is not an even function in $\mu$, and our perturbation ansatz goes as follows ({\it i.e.} perturbative modes of $\alpha (\zeta)$ are not restricted to odd (even) powers of $\ep$ any more).
\beq
\alpha(\mu) &=  \text{sech}\left(\frac{\mu}{L}\right)\sum_{n=1} \alpha_n(\mu)\epsilon^n,
\nn \\
\zeta(\mu) &=\sum_{n=0} \zeta_n (\mu)\epsilon^n ,
 \\
e^{A(\mu)} &= \frac{L}{\ell}\cosh\left(\frac{\mu}{L}\right)\left(1+\sum_{n=1} {\cal A}_n(\mu)\epsilon^n\right).
\nn
\eeq
Note that $W^2$ is ${\cal O}(\ep^2)$, which implies the ${\cal O}(\ep^0)$ part of the equation is exactly the same as the $SO(4) \times SO(4)$ model and we have again 
\beq
\zeta_0 (\mu)&= \zeta_*+ \tan ^{-1}\left(\sinh \left(\frac{\mu}{L}\right)\right)
%,\nn\\ L&= ({\sqrt{2} g})^{-1}
.
\eeq
And of course the equations for $\alpha, A$ are satisfied at $\cO(\ep^0)$ for vacuum configurations.

We find the following solutions for $\cO(\ep)$.
\beq
\alpha_1 (\mu)&=\alpha_{(1)},
\\
\zeta_1(\mu)&=\zeta_{(1)}+\alpha_{(1)}\text{sech}^2\left(\frac{\mu}{L}\right) \left(\sinh \left(\frac{\mu}{L}\right) \cos 3 \zeta_*-\sinh^2\left(\frac{\mu}{L}\right)\sin 3 \zeta_*\right),
\\
{\cal A}_1(\mu)&= -\mu_{(1)}\tanh \left(\frac{\mu}{L}\right).
\eeq
Without losing generality we can set $\alpha_{(1)}=1$, and $\zeta_{(1)}$ can be set to zero since it can be absorbed into re-definition of $\zeta_*$. And $\mu_{(1)}$ can be also set to zero since it corresponds to the 
translational freedom in $\mu$. 

Substituting the $\cO(\ep)$ results into $\cO(\ep^2)$ equations and proceeding in the same way, we have 
\beq
\alpha_2(\mu)&=\text{sech}^2\left(\frac{\mu}{L}\right) \left(\sinh \left(\frac{\mu}{L}\right) \sin 3 \zeta_*- \cos 3 \zeta_*\right),
\\
\zeta_2(\mu)&= \tanh ^4\left(\frac{\mu}{L}\right) \sin 6 \zeta_*+\frac{1}{4} \left(1-3 \cosh \left(\frac{2\mu}{L}\right)\right) \tanh \left(\frac{\mu}{L}\right) \text{sech}^3\left(\frac{\mu}{L}\right) \cos 6 \zeta_*,
\\
{\cal A}_2(\mu)&= -\frac{3}{2} . 
\eeq
We again made use of the re-definition freedom of $\ep$, and 
the integral constant for ${\cal A}_2$ is fixed as we require $A'=0$ at $\mu =0$.

At third order in $\ep$, we find
\beq
\alpha_3(\mu)&= \frac{1}{48} \text{sech}^4\left(\frac{\mu}{L}\right) \left[6 \left(\sinh \left(\frac{\mu}{L}\right)-3 \sinh \left(\frac{3\mu}{L}\right)\right) \sin 6 \zeta_*\right.
\nn\\
&\left.+6 \left(7 \cosh \left(\frac{2\mu}{L}\right)+3\right) \cos 6 \zeta_*-20 \cosh ^2\left(\frac{\mu}{L}\right)\right],
\eeq
\beq
\zeta_3 (\mu) &=\frac{1}{24} \left(\text{sech}\left(\frac{\mu}{L}\right) \left(16 \text{sech}^3\left(\frac{\mu}{L}\right)-45 \text{sech}\left(\frac{\mu}{L}\right)+32\right)-3\right)\sin  3 \zeta_*
\nn\\
&+\frac{1}{24} \left(32 \text{sech}^6\left(\frac{\mu}{L}\right)-96 \text{sech}^4\left(\frac{\mu}{L}\right)+99 \text{sech}^2\left(\frac{\mu}{L}\right)-35\right)\sin9 \zeta_*
\nn\\
&+\frac{1}{24} \tanh \left(\frac{\mu}{L}\right) \text{sech}\left(\frac{\mu}{L}\right) \left(16 \text{sech}^2\left(\frac{\mu}{L}\right)-31\right)\cos3\zeta_*
\nn\\
&+\frac{1}{24} \tanh \left(\frac{\mu}{L}\right) \text{sech}\left(\frac{\mu}{L}\right) \left(32 \text{sech}^4\left(\frac{\mu}{L}\right)-80 \text{sech}^2\left(\frac{\mu}{L}\right)+63\right)\cos9 \zeta_*,
\eeq
\beq
{\cal A}_3(\mu) & = -\frac{4}{3}  \tanh\left(\frac{\mu}{L}\right) \left(\text{sech}^3\left(\frac{\mu}{L}\right)-1\right)\sin3 \zeta_*
\nn\\
&+\frac{1}{6} \left(2 \cosh \left(\frac{2 \mu}{L}\right)+\cosh \left(\frac{4 \mu}{L}\right)+9 \right) \text{sech}^4\left(\frac{\mu}{L}\right)\cos3\zeta_* .
\eeq

We do not present the solutions at higher orders, but obviously it is just the repetition of similar integration problem. 
Using the higher-order results, the Noether charge is found to be 
\beq
\zeta'e^{3A}\sinh^2 2\alpha &=\frac{4 L^2 }{\ell^3}\epsilon ^2-\frac{4  L^2 }{\ell^3}\cos (3 \zeta_*)\epsilon ^3+\frac{6 L^2}{\ell^3}  (\cos (6 \zeta_*)-3)\epsilon ^4
\nn\\
&+\frac{L^3 }{6 \ell^3} (64 \sin (3 \zeta_*)+235 \cos (3 \zeta_*)-63 \cos (9 \zeta_*))\epsilon ^5+{\cal O}\left(\epsilon ^6\right).
\eeq
Note that it does depend on $\zeta_*$, the initial condition of $\zeta$. The phase still changes by $\pi$ between $\mu=-\infty$ and $\mu=\infty$, as in the previous 
case of $SO(4)\times SO(4)$.
\beq
\Delta\zeta:=\lim_{\mu \rightarrow \infty} ( \zeta(\mu)-\zeta(-\mu) )=\pi+ {\cal O}(\epsilon^7).
\eeq

One can draw various Janus curves in $(\alpha\cos\zeta,\alpha\sin\zeta)$-plane, and some samples are presented in Fig.\ref{janus_plots}. Since all the solutions flow to $\alpha=0$ as $\mu\ra\pm\infty$, they make a contractible loop. When compared with the plots presented in \cite{Bobev:2013yra}, our perturbative method restricts us to solutions homotopic to AdS vacuum but otherwise we find good agreements. Having $\Delta\zeta=\pi$ for $SO(4)\times SO(4)$ and $SU(3)\times U(1)\times U(1)$ imply that the points at $\mu=\pm\infty$ are smoothly joined at $z=0$. As we will see in the next subsection, it is not the case for $G_2$.
%%%%%%%%%%%%%%%%%
\subsection{The $G_2$ case}
For this truncation, we have $k=7$ and in terms of $\alpha,\zeta$ the (super)-potential is given as follows.
\beq
{\cal P}&= \frac{1}{8} \sinh ^72\alpha \cos 7 \zeta 
\nn\\
&+\frac{1}{32}\cosh ^32\alpha \left(56 \sinh ^42\alpha \cos (4 \zeta )-68 \cosh 4\alpha+25 \cosh (8 \alpha )-149\right)
\nn\\
&+ \frac{7}{16}\sinh ^52\alpha \left(2 (\cosh 4\alpha+3) \cos 3\zeta+(7 \cosh 4\alpha+17) \cos \zeta \right),
\\
{\cal W}&= \sqrt{2} \left( \cosh^7\alpha + 7 \cosh^3 \alpha \sinh^4 \alpha e^{4 i \zeta}+ 7 \cosh^4 \alpha \sinh^3 \alpha e^{3 i \zeta}+\sinh^7 \alpha e^{7 i \zeta}\right) . 
\eeq
Since the potential has an explicit dependence on the phase $\zeta$, this model does not enjoy a conserved charge, unlike previous examples.

\begin{figure}[t!]
\centering
\includegraphics[width=0.32 \textwidth]{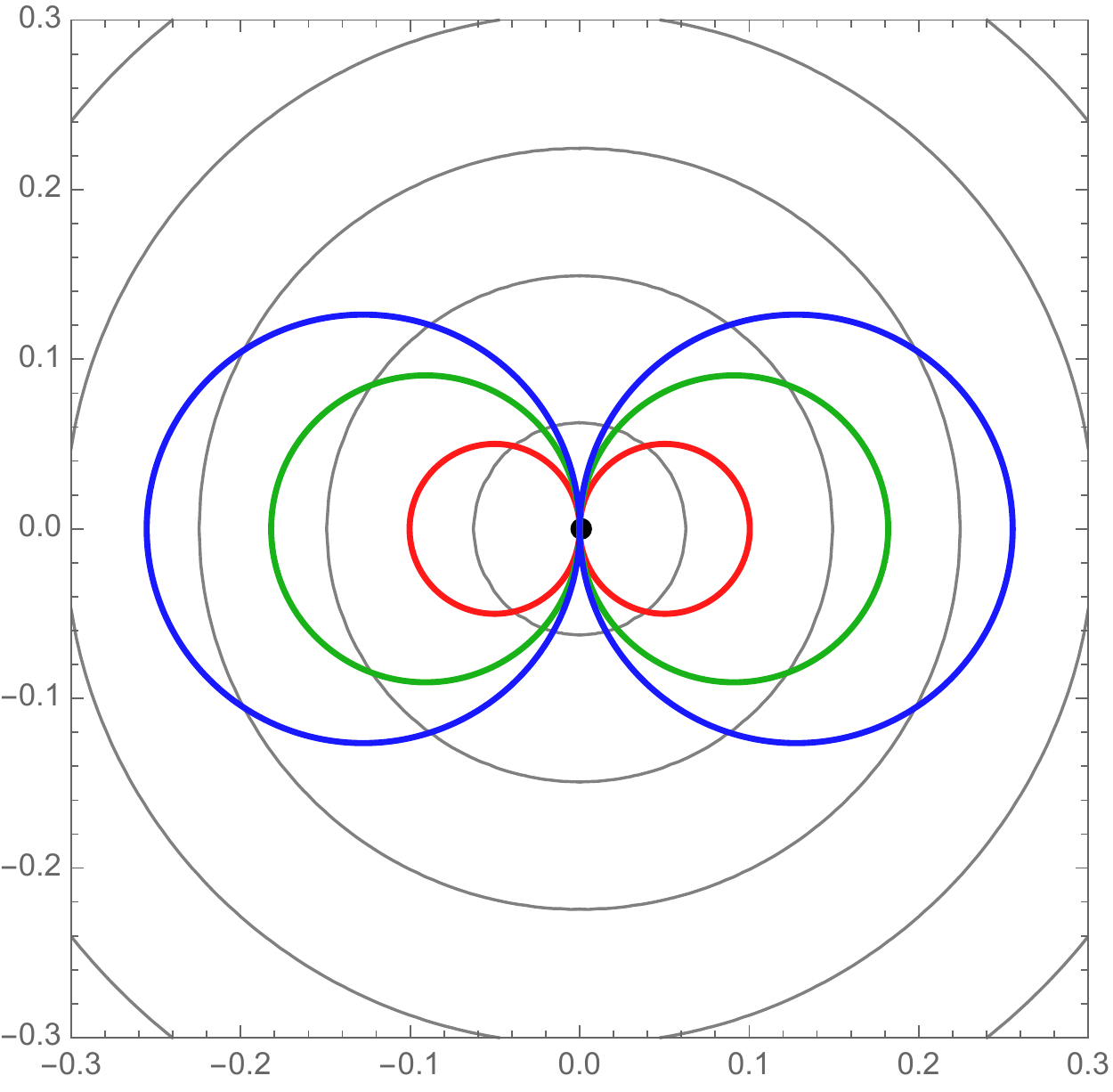} 
\ 
\includegraphics[width=0.32 \textwidth]{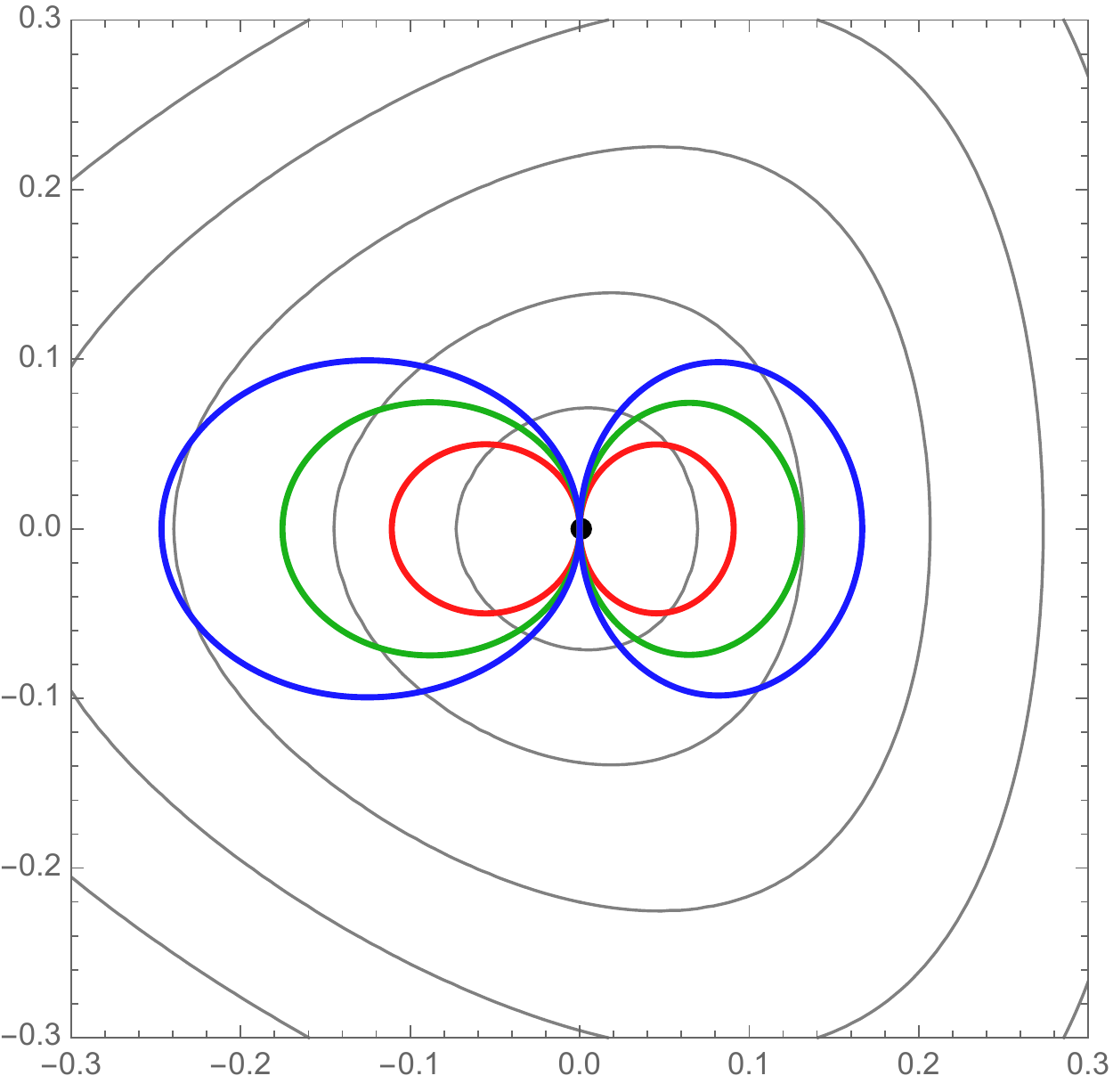} 
\ 
\includegraphics[width=0.32\textwidth]{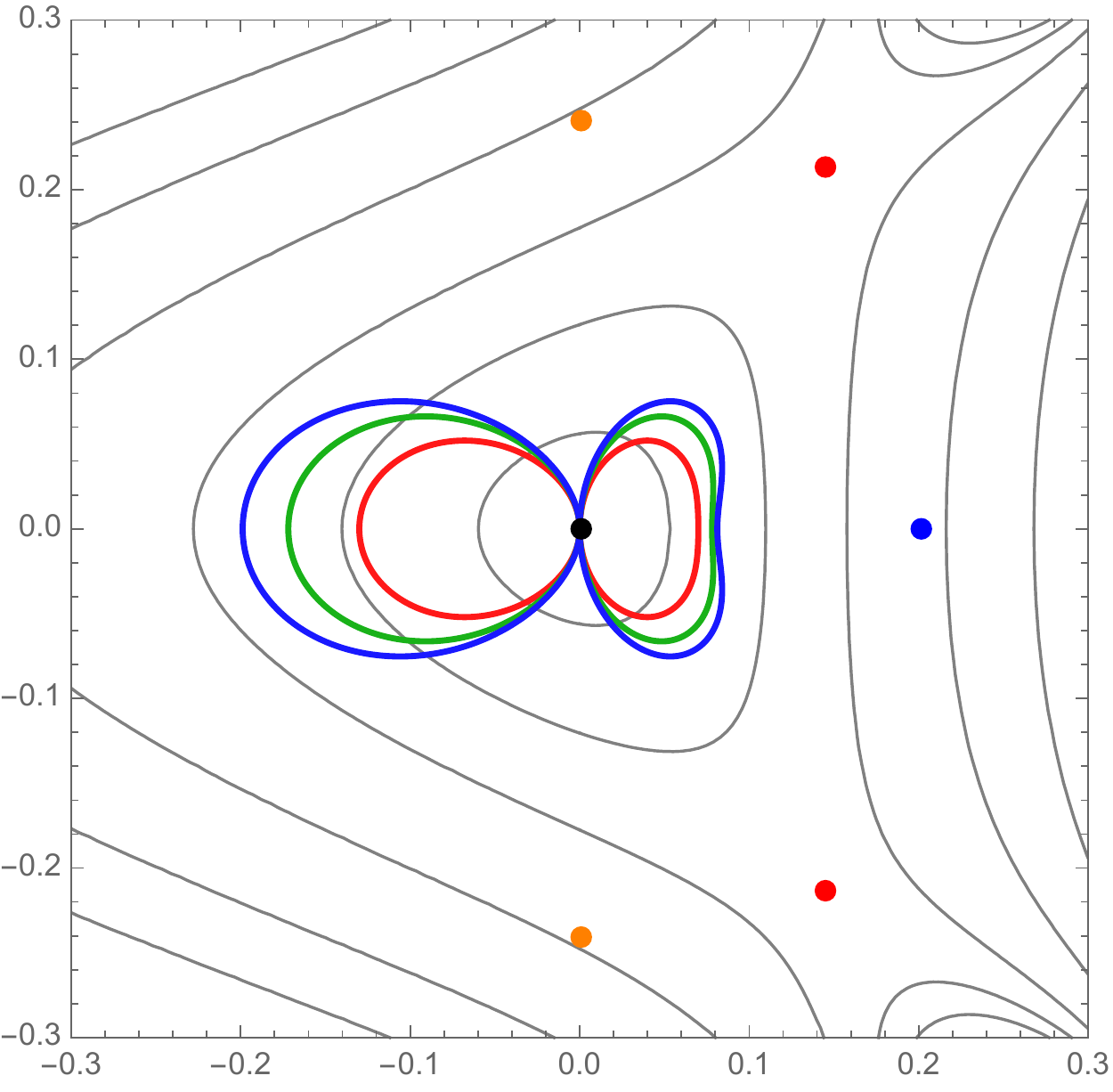} 
\caption{From the left to the right, the figures illustrate Janus solutions in polar coordinates of $\alpha e^{i\zeta}$, for $SO(4)\times SO(4)$, $SU(1) \times U(1) \times U(1)$, and $G_2$-symmetrically truncated models respectively. 
Different colors denote different values of $\ep$, \emph{i.e.}
$0.1, 0.18, 0.25$ for ${SO(4)\times SO(4)}$, $0.1, 0.15, 0.2$ for ${SU(3) \times U(1) \times U(1)}$ and $0.1, 0.125, 0.14$ for $G_2$. Gray lines represent constant-$W$ contours. The maximally supersymmetric vacuum is located at the origin, and on the right panel additional supersymmetric and non-supersymmetric fixed points are also specified in blue, red, and orange colors.
%, and the left and right line in each figure are for $\zeta_*=0, \pi$.
}
\label{janus_plots}
\end{figure}

It is straightforward to write down the BPS equations, and because they are rather lengthy we choose to relegate the 
formulas to Appendix \ref{AppA}. What is important to note is that, unlike previous examples, this model includes five non-trivial $AdS$ fixed points, in addition to the trivial vacuum at $\alpha=0$: There is a non-supersymmetric point with $SO(7)^+$ symmetry (blue dot) at $\alpha = \frac{1}{8} \log 5 $ and $\zeta=0$. Two non-supersymmetric points appear with $SO(7)^-$ symmetry (orange dots) at $\alpha = \frac{1}{2} \text{arccsch} 2$ and $\zeta =\pm \frac{\pi}{2}$. And there are two supersymmetric $G_2$-invariant points, $G_2^\pm$, (red dots) at $\alpha= \frac{1}{2} \text{arcsinh}\left( \sqrt{\frac{2 \sqrt{3}-2}{5}}\right)$ and $\zeta= \pm \text{arccos} \frac{1}{2} \sqrt{3 -\sqrt{3}}$. Their distribution in $z,\bar z$ plane can bee seen in Fig.\ref{janus_plots}.

Although the equations are apparently more complicated, one can proceed perturbatively as with the previous example. The 
zeroth order behavior is the same, and at first order with an appropriate choice of $\ep$, we have
\beq
\alpha _1(\mu)&= 1,
\\
\zeta _1(\mu)&= 3 \text{sech}^2\left(\frac{\mu}{L}\right) \left(\sinh \left(\frac{\mu}{L}\right) \cos 3 \zeta_*-\sinh^2 \left(\frac{\mu}{L}\right)\sin 3 \zeta_*\right),
\\
{\cal A}_1(\mu)&= 0 . 
\eeq
Second order results are as follows.
\beq
\alpha _2(\mu)&= 3 \text{sech}^2\left(\frac{\mu}{L}\right) \left(\sinh \left(\frac{\mu}{L}\right) \sin 3 \zeta_*-\cos 3 \zeta_*\right),
\\
\zeta _2(\mu)&= \frac{1}{8} \tanh \left(\frac{\mu}{L}\right) \text{sech}^3\left(\frac{\mu}{L}\right) \left[72 \sinh ^3\left(\frac{\mu}{L}\right) \sin 6 \zeta_*+18 \left(1-3 \cosh \left(\frac{2\mu}{L}\right)\right) \cos 6 \zeta_* \right.
\nn\\
&\left.-8 \sinh \left(\frac{\mu}{L}\right) \left(\cosh \left(\frac{2\mu}{L}\right)+5\right) \sin 4 \zeta_*+32 \cosh ^2\left(\frac{\mu}{L}\right)+32 \cos 4 \zeta_*\right]
\nn\\
&+3 \pi -12 \tan^{-1} e^{\frac{\mu}{L}},
\\
{\cal A}_2(\mu)&= -\frac{7}{2}.
\eeq
Third order results can be found in the Appendix.

In this model Noether charge theoretically dose not exist, and accordingly $\Delta\zeta\neq\pi$ in general. 
\beq
\Delta\zeta:=\lim_{\mu \rightarrow \infty} ( \zeta(\mu)-\zeta(-\mu) )=\pi-6 \pi \epsilon^2 -15 \pi (\cos (\zeta_*)-3\cos(3\zeta_*))\epsilon^3+{\cal O}(\epsilon^4) .
\eeq
Namely, the two end points $\mu=\pm\infty$ meet at $\alpha=0$ with a cusp.

%%%%%%%%%%%%%%%%%%
\section{\label{sec:4}Entanglement Entropy}
As an application of our perturbative solutions, we will construct minimal area surfaces and the associated 
holographic entanglement entropy \cite{Ryu:2006bv,Ryu:2006ef} from the regularized area via $S_{HEE}=\frac{Area}{4G_N}$. We note that a similar study has appeared
in {\it e.g.} \cite{Estes:2014hka,Gutperle:2015hcv} for (non)-supersymmetric solutions, and one of the authors has considered evaluation of perturbatively obtained time-dependent gravity solutions in \cite{Kim:2015rvu}.

\begin{figure}[h!]
\centering
\includegraphics[width=0.45 \textwidth]{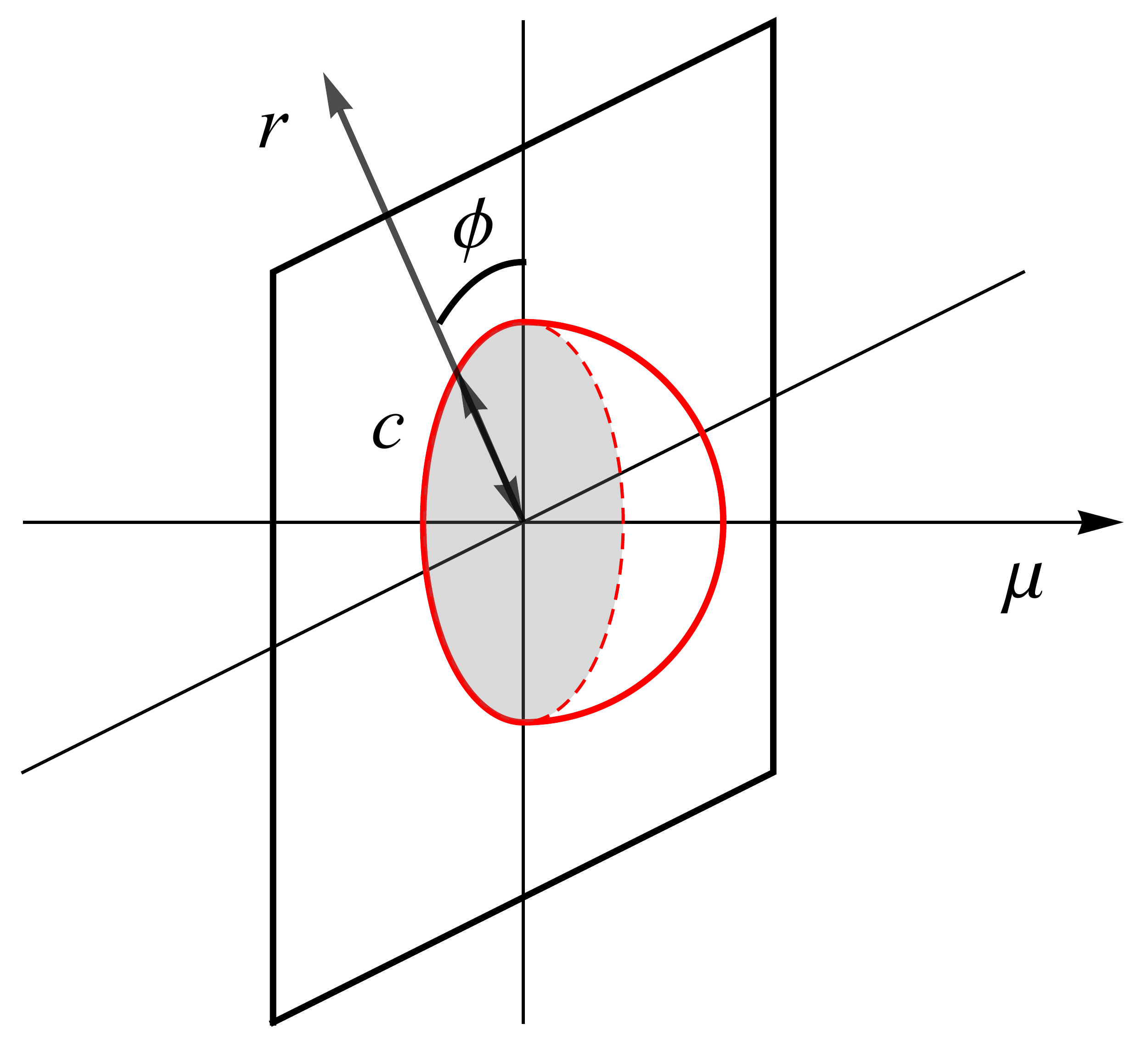} 
\caption{Minimal area as holographic entanglement entropy when the boundary metric is AdS.}
\end{figure}

Our choice for the AdS${}_4$ metric is
\beq
ds^2 &=d\mu^2 +(L/\ell)^2 \cosh^2 (\mu /L)(dr^2 -\cosh^2 (r/\ell) dt^2+ \ell^2\sinh^2 (r / \ell) d\phi ^2),
\eeq
and we choose a disk of radius $r_0$ on the boundary $\mu=0$ and centered at $r=0$ as the entanglement region.
Then the holographic entanglement entropy is given in terms of  
\beq
{\rm Area}&=2\pi  L \int_0^{r_0} dr  \cosh \left(\frac{\mu}{L}\right)\sinh \left(\frac{r}{\ell}\right)\sqrt{\left(\frac{d\mu}{dr}\right)^2 + \left(\frac{L}{\ell}\right)^2\cosh^2\left(\frac{\mu}{L}\right)} .
\label{area4}
\eeq

Through variation one obtains a non-linear 2nd-order differential equation for $\mu(r)$, whose solution can be found thanks to the embedding of AdS${}_4$ inside ${\mathbb R}^{2,3}$ (see Appendix \ref{AppB} for extension to general dimensions).
\beq
\tanh\left(\frac{\mu}{L}\right) = c \cosh \left(\frac{r}{\ell}\right),
\label{minarea}
\eeq
where 
%$c\,\, (0<c<1)$ 
$0<c<1$ and otherwise it 
is an arbitrary constant. It is easy to see that the entanglement region is small (large) when $c\sim 1 \,\,(c\sim 0)$. Substituting the solution into the area \eqref{area4} and introducing a cutoff 
$\delta=(1-\tanh\left(\mu_{max}/L\right))^{1/2}$, one obtains
\beq
\label{heezero}
{\rm Area}&=2\pi L^2\left(\frac{\sqrt{1- c^2} }{\sqrt{2}c  }\frac{1}{\delta}-1-\frac{3 \sqrt{1-c^2}   }{4 \sqrt{2} c }\delta-\frac{5 \sqrt{1-c^2} }{32 \sqrt{2} c }\delta ^3\right)+{\cal O}\left(\delta ^4\right) . 
\eeq
It is well known that the entanglement entropy follows perimeter law for conformal field theories, and we see that it is indeed the case here from the behavior of the divergent part ($\delta^{-1}$) for \emph{small} entanglement region $(c\sim 1)$.

Because we are going to expand around the explicit solution \eqref{minarea}, it will be convenient to switch to new 
variables
\beq
y=\tanh\left(\frac{\mu}{L}\right),
\quad
x= \cosh \left(\frac{r}{\ell}\right). 
\eeq
Their range is by definition $x\geq 1$ and $|y| \leq 1$, and for the solution $y=cx$ their range with cutoff $\delta$ becomes $x\geq (1-\delta^2)/c, |y| \leq 1-\delta^2$.
Our metric ansatz changes to  
\beq
ds^2 &=d\mu^2 + e^{2A} (dr^2-\cosh^2 (r/\ell) dt^2+ \ell^2 \sinh^2 (r / \ell) d\phi ^2),
\nn\\
&=\frac{L^2}{(1-y^2)^2}dy^2 +e^{2A}\left(\frac{\ell^2}{x^2-1}dx^2-x^2 dt^2 +\ell^2(x^2-1)d\phi^2 \right),
\eeq
while the area integral takes the following form.
\beq
{\rm Area}
%&=2\pi \ell  \int_0^{r_0} dr \ e^{A}\sinh \left(\frac{r}{\ell}\right)\sqrt{\left(\frac{d\mu}{dr}\right)^2 + e^{2A}},
%\nn\\
&=2\pi \ell L\int_1^{x_0} dx \frac{e^{A}\sqrt{x^2-1}}{1-y^2} \sqrt{\left(y'\right)^2+\frac{\ell^2(1-y^2)^2}{L^2}\frac{e^{2A}}{x^2-1}} \equiv \int dx {\cal L}(y,y'),
\eeq
where $(\bullet)':=\tfrac{d(\bullet)}{d x}$ in this section. For Janus solutions the metric function $A$ changes, and so does $y(x)$ since the Euler-Lagrange equation changes as well. When we write ${\cal L}={\cal L}_{(0)} + {\cal L}_{(2)} \ep^2+\cdots$ and $y = y_0 + y_2  \ep^2+ \cdots$, using the fact that $y_0$ satisfies the minimal-area condition we find
\beq
{\rm Area} &= \int^{x_0}_1 dx\left( {\cal L} (y_0, y_0')+{\cal L} _{(2)} (y_0, y_0')\epsilon^2\right)+ \left(\frac{\delta {\cal L}}{\delta y'}\right)_{(0)} y_2 \Big| _{x=1}^{x=x_0}\epsilon^2+ {\cal O }(\epsilon^3),
\label{parea}
\eeq
with cutoff $x_0=(1-\delta^2)/c$. The first term is ${\cal O}(\ep^0)$ and the answer is already given in \eqref{heezero}, which exhibits linear divergence. On the other hand, for $y_n\, (n\ge 2)$ we will impose the boundary condition $y_2(x=1/c)=0$ to fix the boundary entangling region, and as the consequence we find that ${\cal O}(\ep^2)$ and subsequent terms are always free from divergence, and starts with a finite term as $\delta \ra 0$.

Let us sketch the computation. We substitute the following expression into the minimal-area condition.
\beq
y(x)&=c x\left(1+ \sum_{n=2} \mathfrak{y}_n(x) \epsilon^n\right).
\eeq
We find 2nd order linear differential equations for $\my_n$. They take the following form in general.
\beq
\hat L \my_n(x) = a_n F(x)+ H_n(x).
\label{eepert}
\eeq
Namely, the homogeneous part is independent of $n$, and the inhomogeneous part is written as the sum of $n$-independent and universal part $F(x)$, and the remaining part $H_n$ which does depend on $n$. The coefficients $a_n$
is constant, $a_2= -\frac{k}{2}$ and $a_3 = \frac{3}{2}{\tilde k} \cos 3 \zeta_*$ where $k=1, 3, 7$ and $\tilde k=0, 1, 7$ for $SO(4) \times SO(4) $, $SU(3)\times U(1) \times U(1)$ and $G_2$ symmetric models respectively. The differential operator $\hat L$ and the universal part are in fact the same for all three models,
\beq
\hat L f &=\frac{\sqrt{1-c^2x^2}}{x^2(1-x^2)}\frac{d}{dx}\left(\frac{x^2 \left(1-x^2\right)}{\sqrt{1-c^2 x^2}}\frac{df}{dx}\right),
\nn\\
F(x)&=\frac{2c  \left(c^2 \left(x^2+1\right)-2\right)}{\left(x^2-1\right) \left(c^2 x^2-1\right)} .
\eeq
And the model-dependent part $H_n$ for $n=2,3$ are given as 
\beq
H_2(x) & = 0 , 
\nn\\
H_3(x)&=\tilde{k}\Big[\frac{c  \left(8 c^6 x^4 \left(3 x^2-5\right)-24 c^4 x^2 \left(x^2-3\right)-c^2 \left(17 x^2+33\right)+18\right)}{3 \left(x^2-1\right) \left(c^2 x^2-1\right)}\cos(3\zeta_*)
\nn\\
&+\frac{4 \left(c^4 x^2 \left(3 x^2-1\right)-3 c^2 \left(x^2-1\right)-2\right)}{3 x \left(x^2-1\right) \left(c^2 x^2-1\right)}\sin( 3 \zeta_*)
\nn\\
&+\frac{4  \left(2 c^4 x^2 \left(3 x^2-5\right)+3 c^2 \left(x^2+1\right)-2\right)}{3 x \left(x^2-1\right)}\sqrt{1-c^2 x^2}  \sin( 3 \zeta_*) \Big] . 
\eeq
Readers might wonder why the inhomogeneous part, {\it i.e.} the right-hand-side of \eqref{eepert} takes similar forms
for different models. It is because the warp factor $e^A$ takes the following universal form, at least up to $\ep^3$. 
\beq
e^{A}&=\frac{L}{l \sqrt{1-y^2}}\left(1-\frac{k}{2}\epsilon^2+\tilde{k}{\mathscr A}_3\epsilon ^3+{\cal O }(\epsilon^4)\right),
\eeq
where $k=1, 3, 7$ and $\tilde{k}=0, 1, 7$ for $SO(4) \times SO(4) $, $SU(3)\times U(1) \times U(1)$ and $G_2$
respectively, and
\beq
{\mathscr{A}}_3 &=\frac{2}{3}  \left(2y(1-( 1-y^2)^\frac{3}{2} )\sin 3 \zeta_*+ (3-3y^2+2y^4)\cos 3 \zeta_* \right).
\eeq
They begin to differ at $\ep^4$, but for simplicity we consider the minimal surface only up to $\ep^4$ here. 

It is obvious from the form of $\hat L$ that \eqref{eepert} can be treated as a 1st order differential equation for $\mathfrak{y}_n$. Thanks to linearity, the solutions in general take the following form,
\beq
\my_n(x) = \my^{(h)} (x) + a_n \my^{(u)} (x) +\my_n ^{(m)}(x),
\label{eefull}
\eeq
where $\my^{(h)}$ is the homogeneous solution, and $\my^{(u)},\my^{(m)}_n$ are particular solutions for $F, H_n$ respectively.
Homogeneous solutions are easily found,
\beq
\my^{(h)}(x) = c_1 \left(-\frac{\sqrt{1-c^2 x^2}}{x} + \sqrt{1-c^2} \sinh^{-1} \left( \frac{\sqrt{c^2-1} x}{\sqrt{x^2-1}}\right) \right) + c_2.
\eeq
We note that the part with $c_1$ is divergent at $x=1$. Without losing generality we can choose $\my^{(h)}(x)=0$ in 
\eqref{eefull} since it can be included in the in-homogeneous solutions.
We then need to construct particular solutions $\my^{(u)},\my^{(m)}_n$ and impose $\my(1/c)=0$ and regularity at $x=1$.

\begin{figure}[h!]
\centering
\includegraphics[width=0.31 \textwidth]{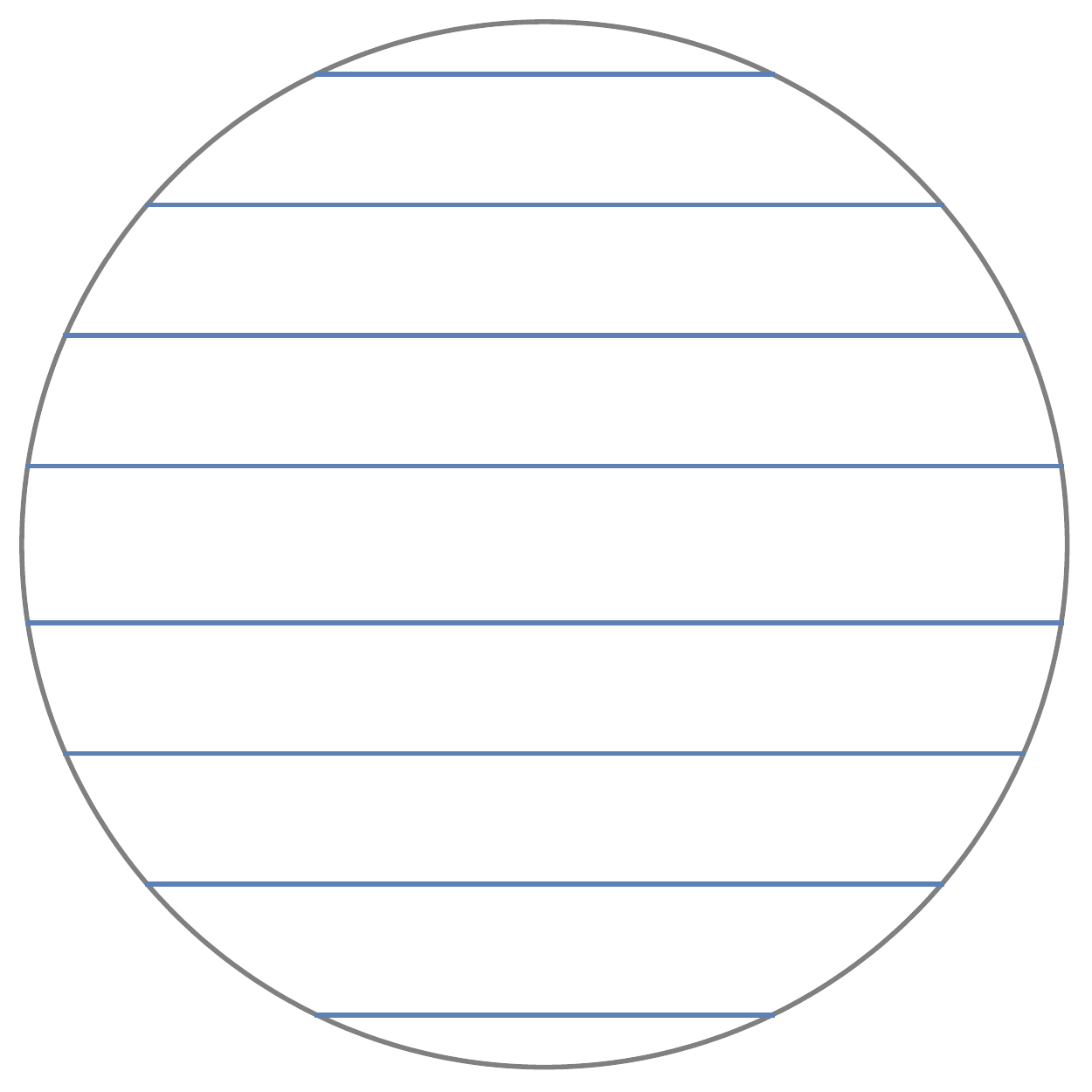} 
\hspace{20pt}
\includegraphics[width=0.31 \textwidth]{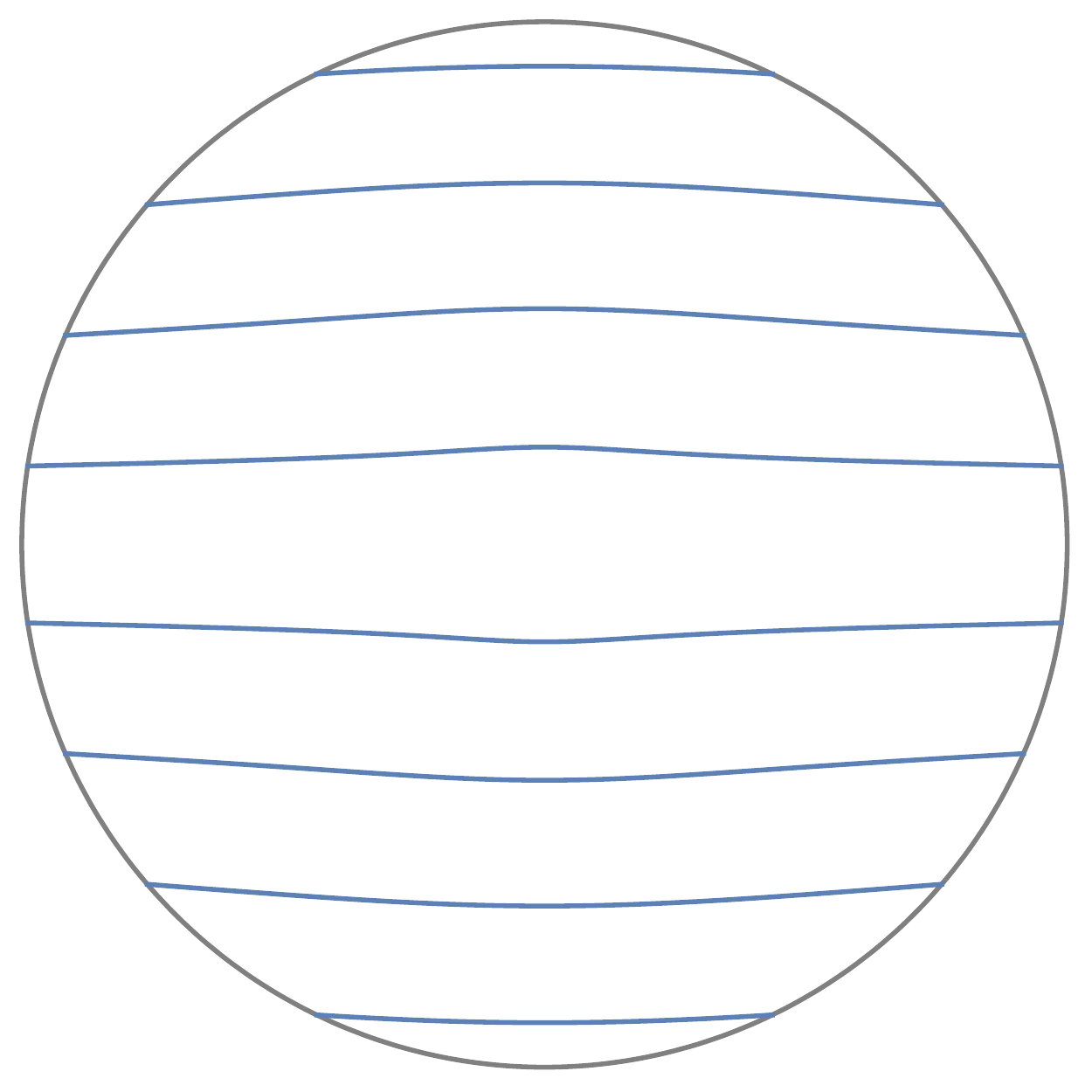} 
\caption{Minimal area surfaces. The left panel is for AdS vacuum and the right panel is for nontrivial Janus backgrounds.}
\label{minsurface}
\end{figure}

Now let us turn to the inhomogeneous part. 
$d(\my^{(u)})/dx$, the first derivate of an inhomogeneous solution for $F(x)$, is given as 
\beq
\frac{d\my^{(u)}}{dx}=\frac{ c^2 \left(x^2-2\right)+1}{c x \left(x^2-1\right)}-\frac{ \sqrt{1-c^2 x^2} \left(c \sqrt{1-c^2}+\left(1-2 c^2\right)( \sin ^{-1}c x- \sin ^{-1}c)\right)}{c^2 x^2 \left(x^2-1\right)}.
\eeq
Unfortunately its integration cannot be done analytically, and we have instead
\begin{footnotesize}
\beq
\my^{(u)}(x) &= \my^{(u)}_* - \frac{\sqrt{1-c^2} \sqrt{1-c^2 x^2}}{c x}+ \frac{1-c^2}{c} \log \left( \sqrt{1-c^2x^2} + \sqrt{1-c^2} x \right) -\frac{1-2c^2}{c} \log x
\nn\\
&-\frac{1-2 c^2} {c^2 x (x^2-1) } \left(\sin^{-1}c - \sin^{-1} cx \right) \left( \sqrt{1-c^2x^2} + c x \sin ^{-1} cx \right)
\nn\\
&-\int_{1/c}^x dx' \frac{(1-2c^2)(\sqrt{1-c^2 x'^2} + cx' \sin^{-1} c x' )} {c^2 x' (x'^2 -1) ^2 } \left(\frac{c (x'^2-1)}{\sqrt{1-c^2x'^2}} + 2 x' (\sin^{-1} c - \sin^{-1} c x' ) \right) ,
\eeq
%\end{small}
\end{footnotesize}
which is finite at $x=1$ and the integration constant $\my^{(u)}_*$ is chosen to guarantee $\my^{(u)}(1/c)=0$.
\beq
\my^{(u)}_* = {c} \log c- \frac{1-c^2}{2c} \log \left( {1-c^2}\right) + \frac{\pi c (1-2c^2) ( \pi - 2 \sin^{-1} c ) }{4 (c^2-1)}.
\eeq
The particular solution due to $H_3$ can be also obtained in the same fashion. Let us just present the result here.
\beq
\my_3^{(m)} &= \tilde{k}\left[\frac{3}{2} \cos 3 \zeta_* \ \my^{(u)}(x) + S(x) \sin3 \zeta_*+ C(x) \cos3 \zeta_* \right],
%\eeq
%\beq
\\
S(x)&=  c ( 1-c^2)^2( 2\sin ^{-1}(c x)- \pi) -\frac{4}{3 x}\left( 1- c^2 x^2\right)
\nn\\
&+\sqrt{1-c^2 x^2} \left(\frac{c^2}{3} \left(c^2 x^3-3 c^2 x+x\right)+\frac{6c^4- 9 c^2 +4}{3 x}\right),
\\
%&+2 c \left(1-c^2\right)^2  \sin ^{-1}(c x)
%\eeq
%\beq
C(x)&=\frac{1}{6 x}c  \left(2 c^4 x^5-12 \left(1-c^2\right)^{3/2} \sqrt{1-c^2 x^2}+\left(6 c^2-3\right) x+\left(c^2-6 c^4\right) x^3\right)
\nn\\
&-c \left(1-c^2\right)^2  \left(2 \log \left(\sqrt{1-c^2} x-\sqrt{1-c^2 x^2}\right)-\log \left(1-c^2\right)+2 \log (c)\right).
\eeq
One can plot the minimal area surface for pure AdS and Janus solutions, see Fig.\ref{minsurface}. Intuitively the minimal surface should be 
pushed away from the center of AdS due to the redshift effect when there is nontrivial excitation. 

We now substitute these solutions into \eqref{parea} and evaluate up to $\ep^3$ terms,
\beq
{\rm Area} = {\rm Area}_{(0)} +{\rm Area}_{(2)} \epsilon^2 + {\rm Area}_{(3)} \epsilon^3 + {\cal O} (\epsilon^4).
\eeq
It turns out that the leading terms of the integral can be evaluated using the differential equation for $\my_n$. 
The results are 
%\beq
%d{\rm Area}_{(2)} =\frac{\pi k L^2 \left(c^2 \left(x^2+1\right)-2\right)}{\sqrt{1-c^2}  \left(1-c^2 x^2\right)^{3/2}}+\frac{d}{d x} \left(\frac{2\pi c L^2 x \left(x^2-1\right) y_2(x)}{\sqrt{1-c^2}  \left(1-c^2 x^2\right)^{3/2}}\right)
%\eeq
\beq
{\rm Area}_{(2)}&= 
\pi L^2 k\left(\frac{ \left(2 c \sqrt{1-c^2} +(\pi-2) \left(c^2-2\right) \sin ^{-1}c \right)}{3 c \sqrt{1-c^2} }+\frac{\sqrt{2-2 c^2}   }{c }\delta\right)+\cO\left(\delta ^2\right) ,
\\
{\rm Area}_{(3)}&=
2\pi L^2 \tilde{k} \left[\left(\frac{2 \left(c^4+c^2-2\right) \sqrt{1-c^2} }{3 c}+\frac{4 \left(2 c^4+c^2+12\right) \sqrt{1-c^2}  }{15 c}\delta ^2\right) \sin3 \zeta_*\right.
\nn\\
&\left.+\left(\frac{1}{3} \left(1-c^2\right) \left(2 c^2-3\right) +\frac{3 \sqrt{1-c^2} }{\sqrt{2} c}\delta-\frac{8}{15} c^4 \delta ^2\right)\cos3\zeta_*\right]
\nn\\
&-3\frac{\tilde{k}}{k} \cos 3 \zeta_* \ {\rm Area}_{(2)}+\cO\left(\delta ^3\right).
\eeq

\section{\label{sec:5}Discussions}
In this paper we have applied a perturbative technique, where we expand the supergravity equations
around a pure AdS configuration in the expansion 
parameter which is one of the integration constants, and solve the linearized equations order-by-order iteratively. 
We have intended to be illustrative, and 
considered three simple models which are consistent truncations of $D=4, SO(8)$ gauged supergravity
and have studied Janus solutions. Let us stress here that our method is different from the conventional series 
expansion of the field equations near UV ({\it i.e.} near the boundary of AdS), where the IR boundary 
condition cannot be incorporated analytically and one usually has to rely on numerical integration. In our method we
instead impose the IR boundary condition at every order in $\ep$, and the holographically renormalized quantities
can be exactly obtained as a function of CFT deformation parameters. Although we have considered {\it 
single-scalar} models in this paper, the advantage of our method stands out more strongly when we consider
multi-scalar models (see {\it e.g.} \cite{Kim:2019rwd}) where thorough numerical analysis is much more time-consuming. 

There are obviously several avenues to investigate further. One is to study other supergravity 
models. There are many works on supersymmetric Janus solutions in various dimensions 
\cite{DeWolfe:2001pq,Clark:2004sb,Clark:2005te,DHoker:2006vfr,DHoker:2007zhm,Suh:2011xc,Bachas:2013vza,Janik:2015oja,Pilch:2015dwa,Bak:2016rpn,Karndumri:2016tpf,Karndumri:2017bqi,Gutperle:2017nwo,Suh:2018nmp,Gutperle:2018fea}
and one can obviously apply our method and construct the solutions in a semi-analytic form.

It will be also worthwhile to try to extract other physical quantities from Janus configurations so that one
can compare with the corresponding field theory side computations eventually. We note that in \cite{Korovin:2013gha}
single-scalar models were studied using a first-order formalism, inspired by Hamilton-Jacobi theory, and then the 
result was used to calculated holographic entanglement entropy and boundary OPE. 
We also note that 
the contribution of the interface to the correlation functions and sphere partition functions are discussed in \cite{Melby-Thompson:2017aip}, where the solutions connect two different conformal fixed points. Perhaps from a more fundamental 
perspective, one would like to identify the conformal field theory living on the interface, namely the conformal field 
theory dual of the AdS${}_3$ slice in our setting, from the holographic results of correlation functions, partition function
and entanglement entropy. See {\it e.g.} \cite{Herzog:2019bom} for discussion of how marginal deformation affects the
partition function when the spacetime has a boundary (interface), from the calculations in free field theory.

Let us also point out that there exists an interesting generalization of Janus configurations in the literature. 
One can consider space-modulated deformations, and with an ingenious choice of the ansatz one still obtains ordinary differential equations \cite{Donos:2013eha}, allowing analytic control than the most general
cases where one has to solve partial differential equations. An interesting physical consequence is so-called
boomerang RG \cite{Donos:2017ljs,Donos:2017sba}, 
namely one can avoid analogues of c-theorem and the at both ends of the renormalization group one 
encounters the same conformal field theory. Let us comment that for ABJM model, spatially modulated mass deformations 
were studied both holographically and on the field theory side in a number of papers 
\cite{Kim:2018qle,Gauntlett:2018vhk,Arav:2018njv,Kim:2019kns,Ahn:2019pqy}. One can certainly re-visit the holography
side analysis employing our method, and also study spatially modulated solutions in other AdS/CFT examples. We plan to 
report on these topics in the near future. 
%%%%%%%%%%%%%%%
\appendix
\section{\label{AppA}Minimal Area Surface of AdS inside Embedding Spacetime}
We present here the minimal area surface solutions inside AdS spacetime of general dimensionality.
In global coordinates the metric of $AdS_{d+1}$ can be written as 
\beq
ds^2= \frac{L^2}{\cos^2 \xi } \left(-dt^2 +d\xi^2 + \sin^2 \xi \,
d\Omega^2_{d-1}\right),
%(d\theta^2 + \sin^2 \theta d \Omega^2_{d-2})),
\label{ads_global}
\eeq
where $d\Omega_{d-1}^2$ denotes the $(d-1)$-dimension sphere with unit radius. This can be derived as induced metric on 
surface defined by 
\beq
\sum_{i=1}^d (X^i)^2 -(X^{d+1})^2-(X^{d+2})^2=-L^2,
\label{ads_def}
\eeq
inside ${\mathbb R}^{d,2}$ with natural flat metric. An explicit parametrization which leads to \eqref{ads_global} is
\beq
X^i &=  L\tan \xi \,Y^i ,
\nn\\
X^{d+1} &= L\sec \xi \sin t ,
\\
X^{d+2}&=  L\sec \xi \cos t, 
\nn
\eeq
where $Y^i$ define the spatial part of the boundary $S^{d-1}$, {\it i.e.} $\sum_i (Y^i)^2 = 1$. The definition \eqref{ads_def}  is also useful to derive the relation $ds^2_{AdS_{d+1}} =d\mu^2+\cosh^2\mu \, ds^2_{AdS_{d}}$: 
one can try $X^1=\sinh\mu$ and $X^i=\cosh\mu {\tilde X}^i \,\, (i=2,\cdots, d+2)$ and make ${\tilde X}^{i}\,\, (i=2,\cdots, d+2)$ define $AdS_{d}$. We also note that an alternative representation of global AdS, 
\beq
ds^2 = d\rho^2 - \cosh^2\rho \, dt^2 + \sinh^2\rho \,d\Omega^2_{d-1},
\eeq
is related to \eqref{ads_global} simply through $\sec \xi = \cosh\rho$, or equivalently $\tan \xi = \sinh\rho$.

Now let us consider holographic entanglement entropy as minimal surface area inside bulk AdS \cite{Ryu:2006bv}. 
We can write 
$d\Omega^2_{d-1}=d\theta^2+\sin^2\theta d\Omega_{d-2}$, and 
for simplicity we choose to divide the boundary into two parts separated by \emph{constant latitude} curve, $\theta=\theta_0$. In terms of $\xi(\theta)$ which describes the shape of the surface in the bulk, the area is 
\beq
{\rm Area}=L^{d-1}{\rm vol} (S^{d-2}) \int_0^{\theta_0} d\theta \frac{(\sin \xi \sin \theta)^{d-2}}{\cos^{d-1}\xi} \sqrt{\left( \frac{d\xi}{d\theta}\right)^2 +\sin^2 \xi} . 
\label{area_formula}
\eeq

One can check that the following relation satisfies the Euler-Lagrangian equation derived from \eqref{area_formula}, for constant $c$.
\beq
\cos \theta \sin \xi = c.
\label{min_area_sol}
\eeq
Obviously this equation is equivalent to \eqref{minarea}, when we identify 
$$\sec\xi=\cosh(\mu/L)\cosh(r/\ell),\quad \tan\xi\cos\theta=\sinh(\mu/L).$$
And from the above parametrization, it is easy to see that this curve is equivalent to the following quadratic equation 
\beq
c^{-2}(X^1)^2 -  
(X^{d+1})^2 - (X^{d+2})^2 = 0 .
\eeq
Or, since we consider spatial surface at a given time defined by $X^{d+1}-\tan t X^{d+2}=0$, \eqref{min_area_sol} is an intersection with a plane $X^1 \sin t = c X^{d+1}$.

It is now straightforward to substitute the solution \eqref{min_area_sol} into the integral \eqref{area_formula} and calculate the area. The result for general dimensions can be found {\it e.g.} in \cite{Kim:2015rvu}.

%%%%%%%%%%%%%%%%%%%%%%
\begin{acknowledgments}
%This work was partly carried out while Se-Jin Kim was visiting CERN, supported by the CERN-Korea graduate student program. The authors are also partly supported by National Research Foundation (NRF) grant 2019R1A2C2004880. 
This work was supported by a research grant from Kyung Hee University in 2016 (KHU-20160698).
\end{acknowledgments}
%%%%%%%%%%%%%%%%
\section{\label{AppB}BPS equations for the $G_2$ symmetric truncation}
In this appendix we present the BPS equations and their solutions obtained using our perturbative prescription for
$G_2$ symmetric model. In terms of the superpotential the BPS equation is given as 
\beq
\alpha' &= -\frac{1}{14} \left(\frac{A'}{W^2}\right)\frac{\partial W^2}{\partial \alpha}- \left(\frac{e^{-A}}{7\ell}\right) \frac{1}{\sinh (2 \alpha)}\frac{1}{W^2} \frac{\partial W ^2} {\partial \zeta},
\\
\zeta'&= -\frac{2}{7} \left(\frac{A'}{W^2}\right)\frac{1}{\sinh^2 (2\alpha)}\frac{\partial W^2}{\partial \zeta}+\left(\frac{e^{-A}}{7\ell}\right) \frac{1}{\sinh (2 \alpha)}\frac{1}{W^2} \frac{\partial W^2 } {\partial \alpha},
\eeq
More concretely, one obtains
\beq
\alpha' &= -\frac{1}{64} \left(\frac{A'}{W^2}\right) \sinh 2 \alpha  \left(\sinh 2 \alpha  \cos \zeta +\cosh 2 \alpha \right)^2\times
\nn\\
& \left(-2 \sinh 4 \alpha  \left(4 \sinh 4 \alpha  \cos 4 \zeta +25 \cos \zeta -14 \cos 3 \zeta \right) +\sinh 8 \alpha  \left(10 \cos 3 \zeta -7 \cos \zeta \right)\right.
\nn\\
&\left. +8 \sinh ^3 2 \alpha  \left(\cosh 2 \alpha  \cos 5 \zeta -6 \sinh 2 \alpha  \cos 2 \zeta \right)+8 \cosh 4 \alpha +14 \cosh 8 \alpha +42 \right)
\nn\\
&+ \left(\frac{e^{-A}}{8\ell}\right) \frac{1}{\sinh 2 \alpha}\frac{1}{W^2} \sinh ^3 2 \alpha  \sin \zeta  \left(\sinh 2 \alpha  \cos \zeta+\cosh 2 \alpha \right)^2 \times
\nn\\
& \left(2 \sinh ^2 2 \alpha  \cos 4 \zeta -4 \sinh 4 \alpha  \left(4 \cos \zeta +\cos 3 \zeta \right)+\cosh 4 \alpha  \left(11 \cos 2 \zeta +10\right)+13 \cos 2 \zeta +2\right),
\eeq
\beq
\zeta'&= \frac{1}{4} \left(\frac{A'}{W^2}\right)\frac{1}{\sinh^2 2\alpha}  \sinh ^3 2 \alpha  \sin \zeta  \left(\sinh 2 \alpha  \cos \zeta+\cosh 2 \alpha \right)^2 \times
\nn\\
& \left(2 \sinh ^2 2 \alpha  \cos 4 \zeta -4 \sinh 4 \alpha  \left(4 \cos \zeta +\cos 3 \zeta \right)+\cosh 4 \alpha  \left(11 \cos 2 \zeta +10\right)+13 \cos 2 \zeta +2\right)
\nn\\
&+\left(\frac{e^{-A}}{32 \ell}\right) \frac{1}{\sinh 2 \alpha}\frac{1}{W^2}  \sinh 2 \alpha  \left(\sinh 2 \alpha  \cos \zeta +\cosh 2 \alpha \right)^2\times
\nn\\
& \left(-2 \sinh 4 \alpha  \left(4 \sinh 4 \alpha  \cos 4 \zeta +25 \cos \zeta -14 \cos 3 \zeta \right) +\sinh 8 \alpha  \left(10 \cos 3 \zeta -7 \cos \zeta \right)\right.
\nn\\
&\left. +8 \sinh ^3 2 \alpha  \left(\cosh 2 \alpha  \cos 5 \zeta -6 \sinh 2 \alpha  \cos 2 \zeta \right)+8 \cosh 4 \alpha +14 \cosh 8 \alpha +42 \right),
\eeq
The real superpotential is given as follows.
\beq
W^2&=2 \cosh ^{14}\alpha  \Big(14 \tanh ^{11}\alpha  \cos 3 \zeta +14 \tanh ^{10}\alpha  \cos 4 \zeta +2 \tanh ^7\alpha  (49 \cos \zeta +\cos 7 \zeta )
\nn\\
&+14 \tanh ^4\alpha  \cos 4 \zeta +14 \tanh ^3\alpha  \cos 3 \zeta +\tanh ^{14}\alpha +49 \tanh ^8\alpha +49 \tanh ^6\alpha+1\Big).
\eeq
%\beq
%\frac{\partial W^2}{\partial \alpha}&=\frac{7}{32} \sinh 2 \alpha  \left(\sinh 2 \alpha  \cos \zeta +\cosh 2 \alpha \right)^2\times
%\nn\\
%& \left(-2 \sinh 4 \alpha  \left(4 \sinh 4 \alpha  \cos 4 \zeta +25 \cos \zeta -14 \cos 3 \zeta \right) \right.
%\nn\\
%&+\sinh 8 \alpha  \left(10 \cos 3 \zeta -7 \cos \zeta \right)+8 \sinh ^3 2 \alpha  \left(\cosh 2 \alpha  \cos 5 \zeta -6 \sinh 2 \alpha  \cos 2 \zeta \right)
%\nn\\
%&\left. +8 \cosh 4 \alpha +14 \cosh 8 \alpha +42 \right)
%\nn\\
%\frac{\partial W ^2} {\partial \zeta}&=-\frac{7}{8}  \sinh ^3 2 \alpha  \sin \zeta  \left(\sinh 2 \alpha  \cos \zeta+\cosh 2 \alpha \right)^2 \times
%\nn\\
%& \left(2 \sinh ^2 2 \alpha  \cos 4 \zeta -4 \sinh 4 \alpha  \left(4 \cos \zeta +\cos 3 \zeta \right)+\cosh 4 \alpha  \left(11 \cos 2 \zeta +10\right)+13 \cos 2 \zeta +2\right)
%\eeq
The 3rd order solutions are 
\beq
\alpha_3 (\mu)&=\frac{1}{24} \text{sech}^2\left(\frac{\mu}{L}\right) 
\left[3 \text{sech}^2\left(\frac{\mu}{L}\right) \left(32 \sinh \left(\frac{\mu}{L}\right) \sin 4 \zeta_*+9 \left(\sinh \left(\frac{\mu}{L}\right)-3 \sinh \left(\frac{3\mu}{L}\right)\right) \sin 6 \zeta_*\right. \right.
\nn\\
&\left. \left.+8 \left(\cosh \left(\frac{2\mu}{L}\right)-3\right) \cos 4 \zeta_*+9 \left(7 \cosh \left(\frac{2\mu}{L}\right)+3\right) \cos 6 \zeta_*\right)-58\right],
\eeq
\beq
\zeta_3(\mu) & =9 \text{sech}\left(\frac{\mu}{L}\right) \left(12 \tanh \left(\frac{\mu}{L}\right) \sin 3 \zeta_*+\left(5 \cosh \left(\frac{2 \mu}{L}\right)-7\right) \text{sech}\left(\frac{\mu}{L}\right) \cos 3 \zeta_*\right)\tan ^{-1}e^{\frac{\mu}{L}}
\nn\\
&+3 \left(2 \text{sech}^4\left(\frac{\mu}{L}\right)-7 \text{sech}^2\left(\frac{\mu}{L}\right)+5\right)\sin\zeta_*
\nn\\
&+\frac{1}{24} \left(\text{sech}\left(\frac{\mu}{L}\right) \left(736 \text{sech}^3\left(\frac{\mu}{L}\right)-9 \left(72 \pi  \sinh \left(\frac{\mu}{L}\right)+71\right) \text{sech}\left(\frac{\mu}{L}\right)+224\right)-321\right)\sin3 \zeta_*
\nn\\
&+3 \left(8 \text{sech}^6\left(\frac{\mu}{L}\right)-14 \text{sech}^4\left(\frac{\mu}{L}\right)+\text{sech}^2\left(\frac{\mu}{L}\right)+5\right)\sin7 \zeta_*
\nn\\
&-\frac{9}{64}  \left(35 (\cosh \left(\frac{4\mu}{L}\right)+3)-116 \cosh \left(\frac{2\mu}{L}\right)\right) \tanh ^2\left(\frac{\mu}{L}\right) \text{sech}^4\left(\frac{\mu}{L}\right)\sin9\zeta_*
\nn\\
&+\left(\frac{15 \pi }{2}-6 \tanh ^3\left(\frac{\mu}{L}\right) \text{sech}\left(\frac{\mu}{L}\right)-30\tan ^{-1}e^{\frac{\mu}{L}}\right)\cos \zeta_*
\nn\\
&+\left(\frac{9\pi}{2}   \left(6 \text{sech}^2\left(\frac{\mu}{L}\right)-5\right)+\frac{1}{24} \tanh \left(\frac{\mu}{L}\right) \text{sech}\left(\frac{\mu}{L}\right) \left(736 \text{sech}^2\left(\frac{\mu}{L}\right)-397\right)\right)\cos3\zeta_*
\nn\\
&+3 \tanh \left(\frac{\mu}{L}\right) \text{sech}\left(\frac{\mu}{L}\right) \left(8 \text{sech}^4\left(\frac{\mu}{L}\right)-10 \text{sech}^2\left(\frac{\mu}{L}\right)-3\right)\cos7 \zeta_*
\nn\\
&+\frac{9}{8} \tanh \left(\frac{\mu}{L}\right) \text{sech}\left(\frac{\mu}{L}\right) \left(32 \text{sech}^4\left(\frac{\mu}{L}\right)-80 \text{sech}^2\left(\frac{\mu}{L}\right)+63\right)\cos9\zeta_*,
\eeq
\beq
{\cal A}_3(\mu) & = -\frac{28}{3}  \tanh\left(\frac{\mu}{L}\right) \left(\text{sech}^3\left(\frac{\mu}{L}\right)-1\right)\sin3 \zeta_*
\nn\\
&+\frac{7}{6} \left(2 \cosh \left(\frac{2 \mu}{L}\right)+\cosh \left(\frac{4 \mu}{L}\right)+9 \right) \text{sech}^4\left(\frac{\mu}{L}\right)\cos3\zeta_*.
\eeq

\bibliographystyle{jhep}
\bibliography{j4}

\end{document}